\documentstyle[aps,twocolumn,tighten]{revtex}

\begin{document}

\title{Comment on ``Evidence for Narrow Baryon
Resonances in Inelastic $\bbox{pp}$ Scattering"}
\maketitle

In a recent Letter \cite{tat97}, Tatischeff {\it et al.} have claimed
evidence for 3 neutral baryon resonances $N^*$ between the nucleon and
the $\Delta (1232)$.  Two of these have masses $M=1004$ MeV and 1044
MeV below $m_N + m_{\pi}$ and thus their widths ($\Gamma=4$ to 15 MeV)
are radiative.  The third resonance ($M=1094$ MeV) also might have a
radiative decay.  All the resonances have to contribute to Compton
scattering on the nucleon and result in peaks at energies
$E_\gamma=68$, 112, and 169 MeV, respectively, which were never
observed on protons \cite{bar74,fed91,mac95,hal93} or neutrons
\cite{ros90,luc94} loosely bound in the deuteron.  Since constraints of
this type are very sensitive and were not analyzed in the Letter, we
give estimates below.  For other theoretical constraints see Ref.\
\cite{mas90}.

The differential cross section of $\gamma N$ scattering near the
resonance peak must be equally visible at any scattering angle. For
$j=1/2$ it is just isotropic. For $j=3/2$, the angular distribution
typically follows $\frac18(3\cos^2\theta + 7)$ if a dipole ($E1$ or
$M1$) transition dominates, and is rather flat with $\le 25\%$
deviations from an average magnitude.  The differential cross section
averaged over angles and over a center-of-mass energy interval $\Delta
W$ reads
$$
  \langle \frac{d\sigma_{\gamma N}}{d\Omega_{\rm cm}} \rangle =
    \frac{\pi X}{4E_{\gamma\,\rm cm}^2} =  a X \times \cases{
   7.6, &  $M=1004$ MeV \cr
   3.0, &  $M=1044$ MeV \cr
   1.5, &  $M=1094$ MeV }
$$
where $a=10^7$ nb/sr, $X = (j+\frac12) (\Gamma/\Delta W)
{\rm Br}^2_\gamma$, and the radiative branching ${\rm Br}_\gamma =1$
for the first two states.  (Here we assume $\Delta W \gg \Gamma$, which
we show to be a very good approximation.)

The data of Ref.\ \cite{fed91} on $\gamma p$ scattering near $E_\gamma
= 68$ MeV have a scale of $10{-}15$ nb/sr, with variations of at most 3
nb/sr in energy bins of $\Delta W \simeq 5$ MeV.  Therefore, a $p^*$
resonance near 1004 MeV must have $X < 4\cdot 10^{-8}$ and the total
width $\Gamma < 0.2$ eV seven orders of magnitude less than Tatischeff
{\it et al.} have reported.

If we assume $j=1/2$, the interaction leading to the transition $\gamma
N\leftrightarrow N^*$ is dipole M1 (or E1 depending on the parity of
the resonance), $H_{\rm eff}=-e\vec H\cdot\vec\sigma D$.  Here $D$ is a
transition magnetic (or electric) dipole moment and
$e^2/4\pi=\alpha=1/137$.  The radiative width of the $N^*$ then reads
$\Gamma_\gamma = 4\alpha E_{\gamma\,\rm cm}^3 D^2$.  With
$\Gamma_\gamma < 0.2$ eV, the transition dipole moment of $N^*(1004)$
is $D < 1.0\cdot 10^{-3}$ fm, that is at least three orders of
magnitude smaller than the size of the nucleon.  The wave function of
such a resonance would have a very small overlap with the nucleon wave
function, and it would be very difficult to produce $N^*$ with ordinary
beams.

In the same way, data of Ref.\ \cite{mac95} give an upper limit $\Gamma
< 1.6$ eV for the $p^*(1044)$ resonance, and data of Ref.\ \cite{hal93}
give ${\rm Br}^2_\gamma\Gamma < 7$ eV for the $p^*(1094)$.

Information pertaining to neutral states can be obtained, in principle,
via the reaction $\gamma d\to \gamma n p$ in the kinematics of
quasi-free $\gamma n$ scattering.  Rose {\it et al.} \cite{ros90} have
measured the cross section for neutron knockout in inelastic $\gamma d$
scattering, $\gamma(d,\gamma'n)p$, for quasi-free kinematics at
energies around $E_\gamma=110$ MeV (the energy bin was about 40 MeV).
Although the authors did not extract the differential cross section for
$\gamma n$ scattering, they found agreement between the double
differential cross section $d^2\sigma /d\Omega_{\gamma'} d\Omega_n$ and
the theoretical calculation by Levchuk {\it et al.} \cite{lev94},
obtained with the same kinematical conditions.

Since the observed cross section is dominated by the $\gamma n$
subprocess, rescaling arguments can be used to obtain experimental
estimates for the differential cross section of $\gamma n$ scattering.
This leads to the following result:
$$
  \frac{d\sigma_{\gamma n}}{d\Omega_{\rm lab}}
     = \cases{
   2.5 \pm 0.7~{\rm nb/sr}, & $90^\circ$ \cr
   3.2 \pm 0.7~{\rm nb/sr}, & $135^\circ$ }
$$
at $E_\gamma=110$ MeV.  Accordingly, we find $X < 1.5\cdot 10^{-7}$ and
hence $\Gamma < 6$ eV for the $n^*(1044)$ state.

From data on elastic $\gamma d$ scattering at 69 MeV \cite{luc94} one
can find the following bound for the total widths of the $p^*(1004)$
and $n^*(1004)$ states: $\Gamma_p + \Gamma_n \alt 1.5$ eV.

Thus, the states of $M=1004$ and 1044 MeV with the properties given in
Ref.\ \cite{tat97} are completely excluded by Compton scattering data.
The same is valid for the 1094 MeV state unless its branching ratio is
anomalously suppressed in comparison with a typical value of
${\rm Br}_\gamma\sim\alpha$.

It is worth mentioning that a previous search \cite{ram94} for isospin
3/2 resonances in this mass region gave a null result.

This work was supported in part by a U.S. Department of Energy Grant
No. DE-FG02-97ER41038.

\bigskip
\noindent
A.I. L'vov\\
P.N. Lebedev Physical Institute,\\
Leninsky Prospect 53, Moscow 117924, Russia

\bigskip

\noindent
R.L. Workman\\
Department of Physics,\\
Virginia Tech, Blacksburg, VA 24061


\end{document}